\documentclass[twocolumn,preprintnumbers,amsmath,amssymb]{revtex4}
\usepackage{graphicx}% Include figure files
\usepackage{dcolumn}% Align table columns on decimal point
\usepackage{bm}% bold math
\begin{document}
\title{\bf Spiral graphone and one sided fluorographene nano-ribbons}
\author{M. Neek-Amal$^{1}$, J. Beheshtian$^{2}$, F. Shayeganfar$^{3}$,
 S. K. Singh$^{1}$, J. H. Los$^{4}$ and F. M. Peeters }
\affiliation{\\$^1$Departement Fysica, Universiteit Antwerpen,
Groenenborgerlaan 171, B-2020 Antwerpen,
 Belgium.\\
 $^2$ Department of Physics, Shahid Rajaee teacher
Training University, Lavizan, Tehran 16785-136, Iran.
 \\
$^3$ D\'{e}partement de g\'{e}nie physique et Regroupement
qu\'{e}b\'{e}cois sur les mat\'{e}riaux de pointe (RQMP), \'{E}cole
Polytechnique de Montr\'{e}al, C.P. , succ. Centre-Ville,
Montr\'{e}al, Qu\'{e}., Canada H3C 3A7.\\$^4$
 Institute of Physical Chemistry and
Center for Computational Sciences, Johannes Gutenberg University Mainz,
Staudinger Weg 9, D-55128 Mainz, Germany}

\date{\today}

\begin{abstract}
The instability of a free-standing one sided
hydrogenated/fluorinated graphene nano-ribbon, i.e.
graphone/fluorographene, is studied using ab-initio, semiempirical
and large scale molecular dynamics simulations. Free standing
semi-infinite arm-chair like hydrogenated/fluorinated graphene
(AC-GO/AC-GF) and boat like hydrogenated/fluorinated graphene
(B-GO/B-GF) (nano-ribbons which are periodic along the zig-zag
direction) are unstable and spontaneously transform into spiral structures.
We find that rolled, spiral B-GO and B-GF are energetically more favorable
than spiral AC-GO and AC-GF which is opposite to the double sided
flat hydrogenated/fluorinated graphene, i.e. graphane/fluorographene.
We found that the packed, spiral structures exhibit
unexpected localized HOMO-LUMO at the edges with increasing energy
gap during rolling. These rolled hydrocarbon structures are stable
beyond room temperature up to at least $T$=1000\,K.
\end{abstract}


%%%%%%%%%%%%%%%%%%%%%%%%%%%%%%%%%%%%%%%%%%%%%%%%%%%%%%%%%

\maketitle

\section{Introduction}
The discovery of graphene~\cite{geim} has been a driving force for
the scientific community to synthesize and to characterize new
materials with similar morphologies due to their unique
properties~\cite{10,silice,APL2011,novoselov}. Fully double sided
hydrogenated graphene (GE), i.e. graphane (GA), and double sided
fluorographene are quasi two-dimensional lattices of carbon (C)
atoms {\bf ordered into a buckled honey-comb sublattice}, where
each carbon atom is covalently bonded to hydrogen (H) or fluor (F),
respectively,  in an alternating, chair-like
arrangement~\cite{sluiter-sofo,expF}.
The chemisorption of hydrogen and fluor atoms results in an
important reconstruction of the chemical bonds and angles of the
underlying honeycomb lattice~\cite{mehdi}. This transition from
sp$^2$ to sp$^3$ hybridization turns the graphitic C-C bonds into
single bonds by the formation of additional single C-H and C-F
bonds, which change locally the planar shape of graphene into an out
of plane, angstrom scale buckled geometry~\cite{PRB77}.

Experimentally, it has been shown that GA can be obtained reversibly
starting from a pure GE layer in the presence of atomic
hydrogen~\cite{elias}. It has become an interesting material due to
its potential applications in nanoelectronics~\cite{int2}. On the
other hand, the experimentally obtained single-layer fluorographene
exhibits a strong insulating behavior with a room temperature
resistance larger than 1~T$\Omega$, and a high temperature stability
up to 400 $^oC$~\cite{expF}.

Both in experiments~\cite{elias} and in ab-initio calculations,
flat single-sided hydrogenated graphene, graphone (GO), was found to be
unstable~\cite{PRB77,hasan1,ort,Hern,Wen,PRL2003,PRB2009} which was
also demonstrated by phonon band structure calculations. Supporting
graphene on a substrate and hydrogenating was demonstrated to
produce GO in a recent experiment~\cite{GOexp}. Ab-initio
calculations showed that arm-chair (AC)-GA is more stable than boat
like (B)-GA while B-GO was found to be more stable than
AC-GO~\cite{ACSnano}. Ab-initio calculations also showed that
GO exhibits magnetism due to the localized electrons on the carbon
atoms without hydrogens, in contrast to nonmagnetic graphene and
graphane~\cite{nanolett,bandgap}.
\begin{figure}
\begin{center}
\includegraphics[width=0.9\linewidth]{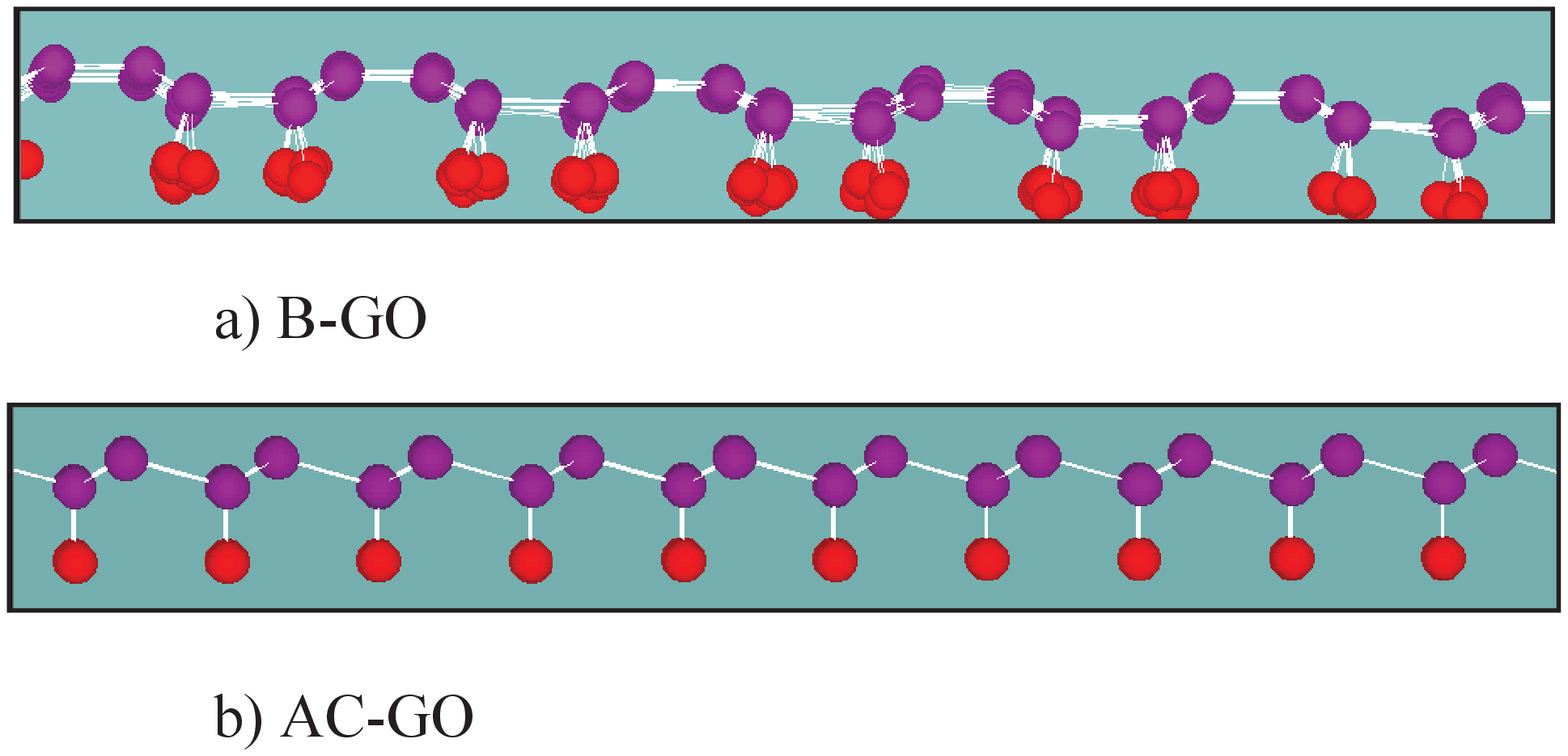}
\caption{(Color online) Side views of a small portion of simulated
graphones which were relaxed at T=10\,K using molecular dynamics
simulation. Arm-chair graphone (a) and boat-like graphone (b).
\label{fig0}}
\end{center}
\end{figure}

By using both ab-initio calculation and elasticity theory, Kudin et
al found that while linear elasticity theory can be used for
studying the stiffness and flexural rigidity moduli of carbon and
boron-nitride nanoshells, one-sided fluorinated carbon ribbons
exhibit a strong tendency to shell formation resulting into very
small diameter tubes with (4,4) and (3,3) indexes and properties
that deviate from linear elasticity theory~\cite{yakobson2001}.
Such a strong tendency to bending is also observed for Si and bilayer SiGe
nanofilms, due to reconstructions in the surface
layer~\cite{PRLSiGe2007} associated with dimer formation. Although
the possible deformation of the edge of graphene
(see~\cite{Ivanovskaya} and references therein) and the synthesis of
carbon nanotubes via one-sided hydrogenation or fluorination of
supported carbon films have been investigated theoretically
\cite{nanolett2007}, the geometrical structure of long (referring to
the distance between the free edges) nano-ribbons of graphone and their
corresponding stability and electronic properties at room temperature
have not been studied so far. Here we show that such long nano-ribbons
of GO/GF do not form carbon nanotubes, as in the case of relatively
short samples with specific sizes~\cite{nanolett2007}, but, instead,
spontaneously form rolled, spiral structures with interesting
localization of frontier molecular orbitals. The one-sided
absorption and corresponding sp$^3$ hybridization is the main
driving force for this transformation, and the final spiral
structure is significantly more stable than the corresponding flat
structures. We used the three mentioned methods to investigate
various electronic and geometrical properties of these new spiral
structures, which we named \emph{spiral graphone (fluorinated
graphene)}'.

There have been several recent experimental and theoretical studies
related to the formation of carbon nanotubes using a graphene sheet
and inverse processing~\cite{prb2012,12012,22012}. Many experimental
evidences point towards the possibility of the production of the
here proposed spiral carbon nanostructures. In a recent review
paper, Vilatela and Eder~\cite{Eder} reviewed related experimental
studies, compared the potentials and characteristics of nano carbon
composites (such as various rolled sheets) and discussed key
challenges for the use of the new carbon nanostructures.
Furthermore, similar rolled structures for Ni$_3$Si$_2$O$_5$(OH)$_4$
were synthesized and analyzed using X-ray diffraction (XRD) and
transmission electron microscopy (TEM)~\cite{PhyChem2011}. Shen et
al~\cite{ACSnano2011} reported recently the synthesis of
self-assembled kinked $In_2O_3$ nanospirals and multikinked
nanowires using a tube-in-tube laser ablation chemical vapor
deposition (CVD) method with gold nanoparticles as the catalysts.

Apart from being more stable than the flat structures, we also show
that spiral B-GO and B-GF are more stable than spiral AC-GO and
AC-GF, respectively. %The absorption pattern (i.e. B or AC) depends
%on the boundary conditions which are controllable and expected to be
%feasible using new nanoscale lithography~\cite{korea} techniques.
The spiral structures are closely packed and stable even at 1000 K
and are candidates for a new class of carbon nano-structures. Since
the deformations for the GFs and the GOs are very similar we will
report here mainly the results for the GOs. After rolling,
the highest occupied orbital (HOMO) and the lowest unoccupied
orbital (LUMO) are separated and appear to be localized at opposite
(free) ends, making these two ends chemically more
active~\cite{fulki}, while before and during rolling they are
localized at both ends simultaneously. We found that the rolling
process makes the systems strongly polarized and more insulating.

\section{Methods and Models}
We performed ab-initio calculations using GAUSSIAN (G09)~\cite{34}
which is an electronic structure package that uses a basis set of
Gaussian type of orbitals. In the ab-inito calculations for the
exchange and correlation (XC) functional, the hybrid functional
B3LYP is adopted in G09. The self consistency loop was iterated
untill the change in the total energy is less than $10^{-7}$ eV, and
the geometries were considered relaxed once the force on each atom
is less than 50\,meV/\AA. Using the 6-31G* basis set in G09, we
expect that our calculation is capable to provide a reliable
description of the electronic properties of the different systems.
In the case of AC-GO the A-sublattice sites of GE were covered with
hydrogens, while the B-GO model was obtained by hydrogenating all carbon
atoms involved in horizontal C-C bonds along the AC direction. Hence
the number of carbon atoms is twice as large as the number of
hydrogen atoms. The simulated samples (using the DFT method) have
typically more than 500 atoms.

For the larger samples, we performed semi-empirical calculations
(PM6 level of calculation in G09) as well as
classical molecular dynamics simulations (MD) using the LAMMPS
package~\cite{lammps}.

For the classical MD simulations, the modified second generation of
Brenner's bond-order potential, i.e. Adaptive Intermolecular
Reactive Bond Order (AIREBO)~\cite{AIREBO} and the
ReaxFF~\cite{REAXFF} potentials were employed to simulate the GOs
and GFs, respectively. These simulations were performed using
periodic boundary conditions along the lateral side (perpendicular
to the rolling direction) in the canonical NVT ensemble with a
Nos\'e-Hoover thermostat for temperature control and a time step of
0.1~fs. The simulation box size along the rolling direction was
taken equal to twice the system size (in its flat geometry) in that
direction.

AIREBO consists of two parts, namely the reactive bond order
potential (REBO \cite{brenner2002}) for the short range interactions
($<$2~\AA), but with a modified, four body, bond order contribution
for the torsion in various hydrocarbon configurations, and the van
der Waals (vdW) term for long range interactions within distances
$2~\AA<r<10~\AA $ which is similar to the standard Lennard Jones
potential, but with an environment dependent (adaptive) suppression
of the (too) strong $ 1/r^{12} $ repulsion. Its total energy reads:

\begin{equation}
E_{AIREBO}=E'_{REBO}+E_{vdW},
\end{equation}
where the prime in $E'_{REBO} $ is added to indicate the
modification in the torsion term. In  our MD simulations we checked
the effect of the vdW term (by turning it on/off) to insure that
when GO is rolled up the vdW term does not prevent possible bonding
between two close edge atoms.  The simulations where we included
the vdW term are labeled with the symbol ${*}$. We found that
including the vdW term lowers the energy of spiral GO* by only
~0.04\,eV/atom with respect to GO.

In ReaxFF (for simulating B/AC-GFs), the atomic interactions are
described by a reactive force field potential~\cite{REAXFF}.
ReaxFF is a general bond-order dependent potential that provides an
accurate description of the bond breaking and bond formation. Recent
simulations on a number of hydrocarbon-oxygen systems showed that
ReaxFF reliably provides energies, transition states, reaction
pathways and reactivity in reasonable agreement with ab-initio
calculations and experiments.

For our MD simulations, our rectangular   GO nano-ribbons contained
different numbers of carbon atoms, namely $N_{C}=4800$ and $5760$.
For the GFs we took $N_{C}=5600$. For the semi-empirical
calculations at the PM3 level, we took $N_C=1500$.

\begin{figure}
\begin{center}
\includegraphics[width=0.49\linewidth]{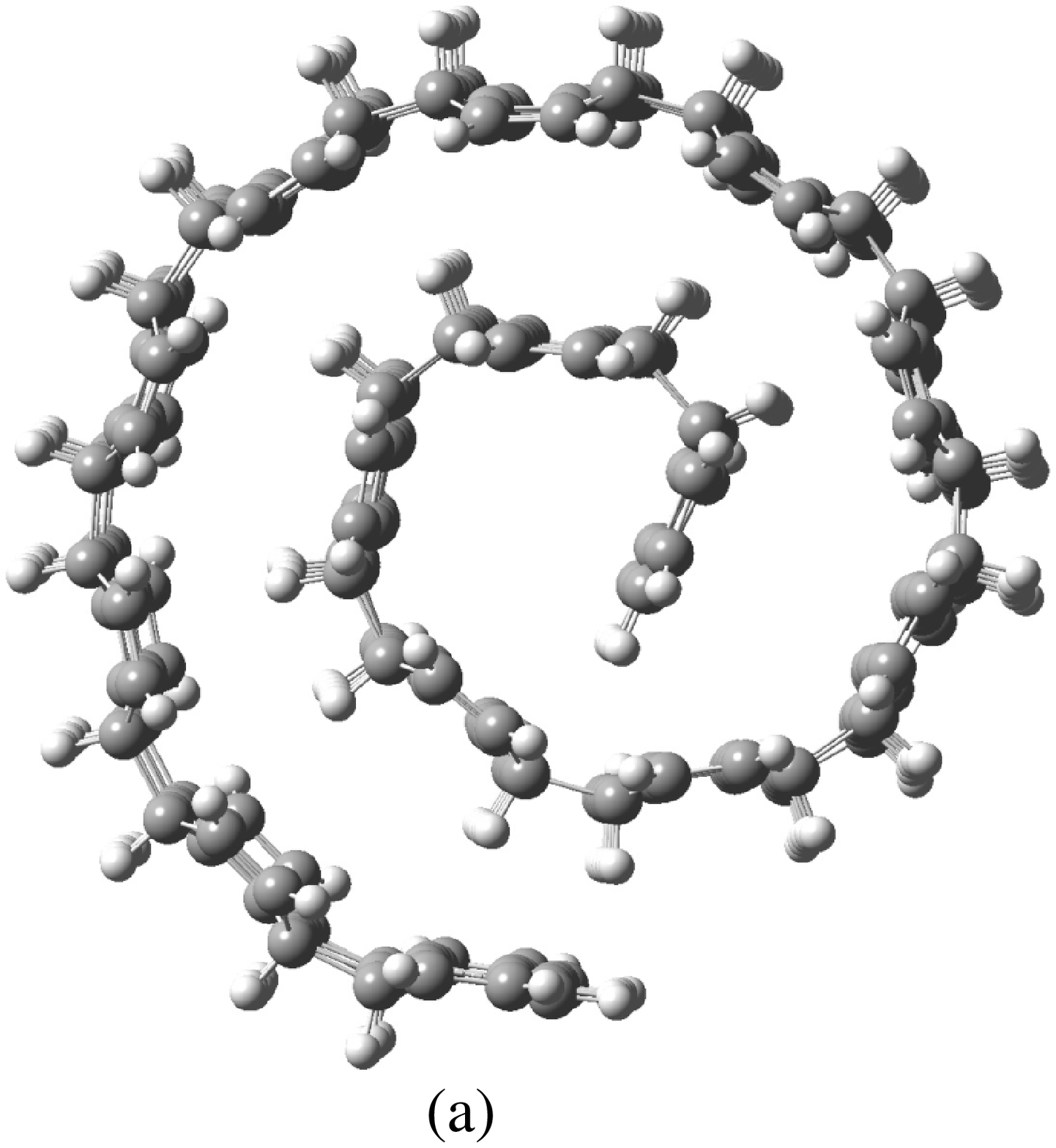}
\includegraphics[width=0.49\linewidth]{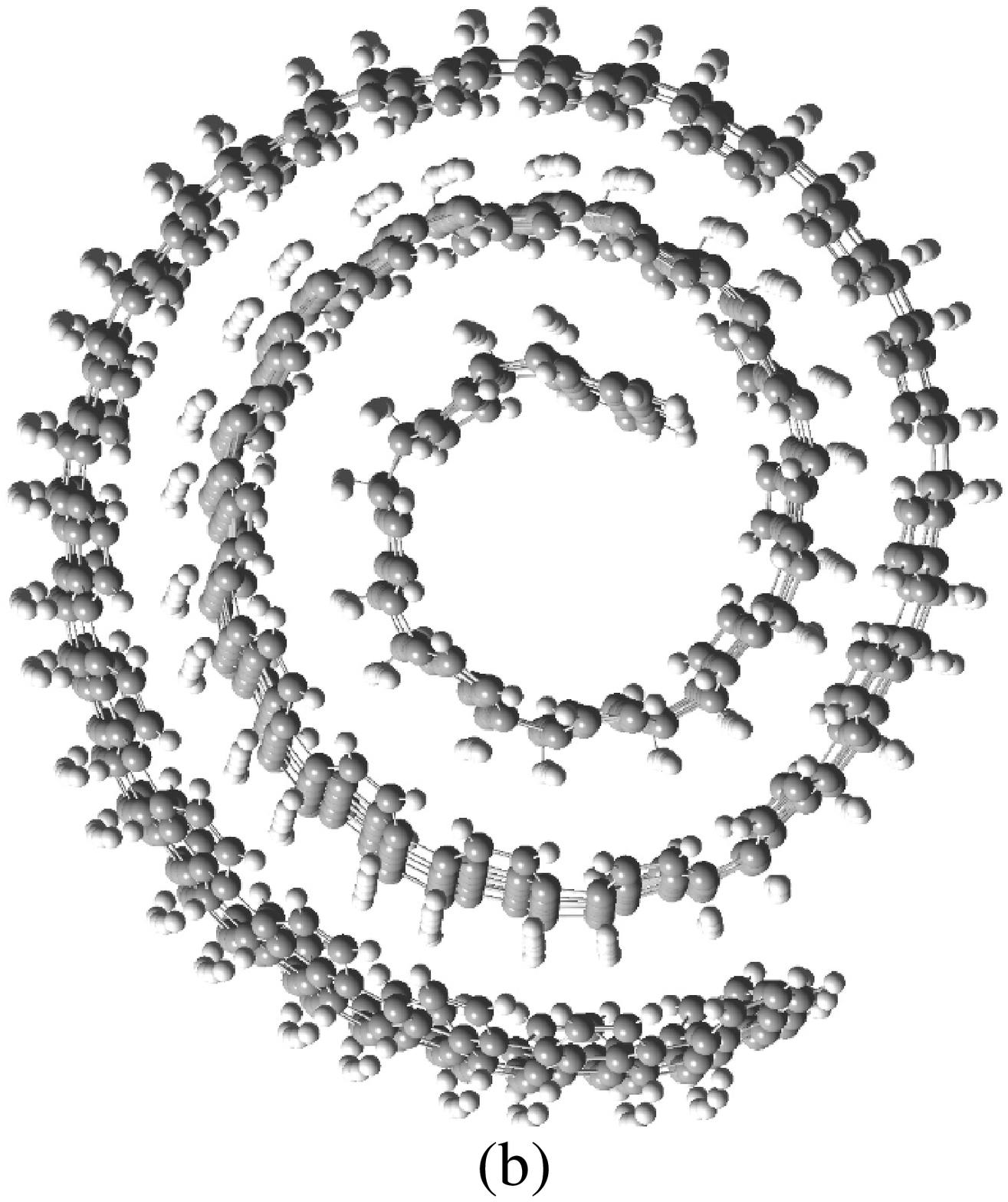}
\caption{(Color online)   (a) The spiral B-GO using DFT calculations
for 504 atoms. (b) The spiral B-GO using semiempirical calculations
(PM6) for 1320 atoms. \label{figDFT}}
\end{center}
\end{figure}

\begin{figure}
\begin{center}
\includegraphics[width=0.95\linewidth]{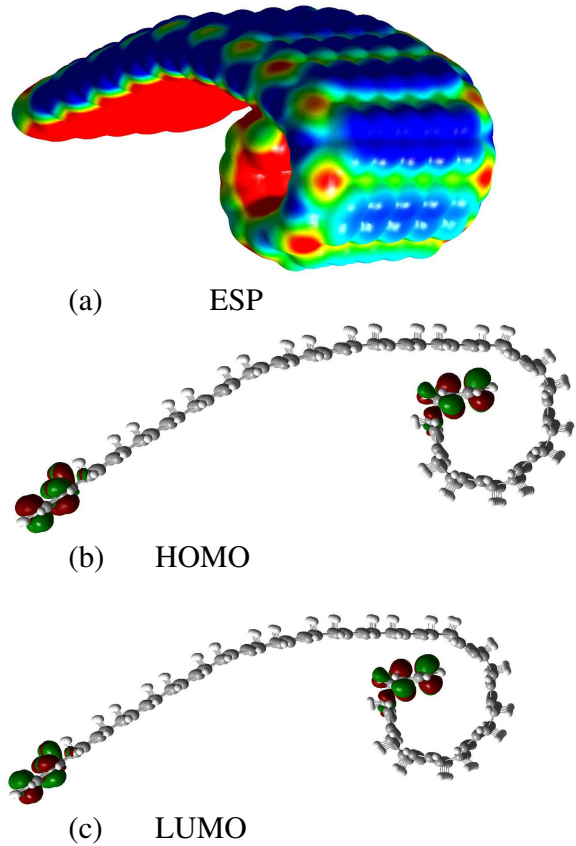}
\caption{(Color online)  Electrostatic potential (a) and the HOMO
(b) and LUMO (c) for B-GO during rolling.\label{figespflat}}
\end{center}
\end{figure}

%%%%%%%%%%%%%%%%%
\section{Results and discussion}
\subsection{Infinite one sided hydrogenated graphene}

Figure~\ref{fig0} shows two side views of relaxed infinite (with
periodic boundary condition applied in both directions of the plane)
B-GO (a) and AC-GO (b) at $T$=10\,K using the MD method. In the
AC-GO configuration, the hydrogens are relatively far from each
other and the C-H bonds are mutually parallel and perpendicular to
the $z=0$ plane, due to symmetry.
In the B-GO the C-H bonds are attached to nearest neighbors C-C bond,
which induces a corrugation in the hydrogen layer with C-H bonds
that are not fully perpendicular to the $z=0$ plane, mainly because
of the sp$^3$ hybridization but also due to the stronger repulsion
between te relatively close hydrogen atoms.
Very similar configurations were obtained for GF, which are therefore
not shown. It is important to note that such flat GOs (and one sided GFs)
are energetically unfavorable~\cite{chemPhysLett2012}. In this study
we report on the structural transformations of these systems into
stable, non-flat structures.

\subsection{Ab-initio results: Graphone}

In Fig.~\ref{figDFT}(a) we show the optimized configuration of B-GO
using  ab-initio calculations with B3LYP/6-31G* and 504 atoms.
Notice that the initial non-relaxed configuration was a flat sheet.
In Fig.~\ref{figDFT}(b) an example of results from our semiempirical
calculation using PM6 is shown.

\begin{figure}
\begin{center}
\includegraphics[width=0.6\linewidth]{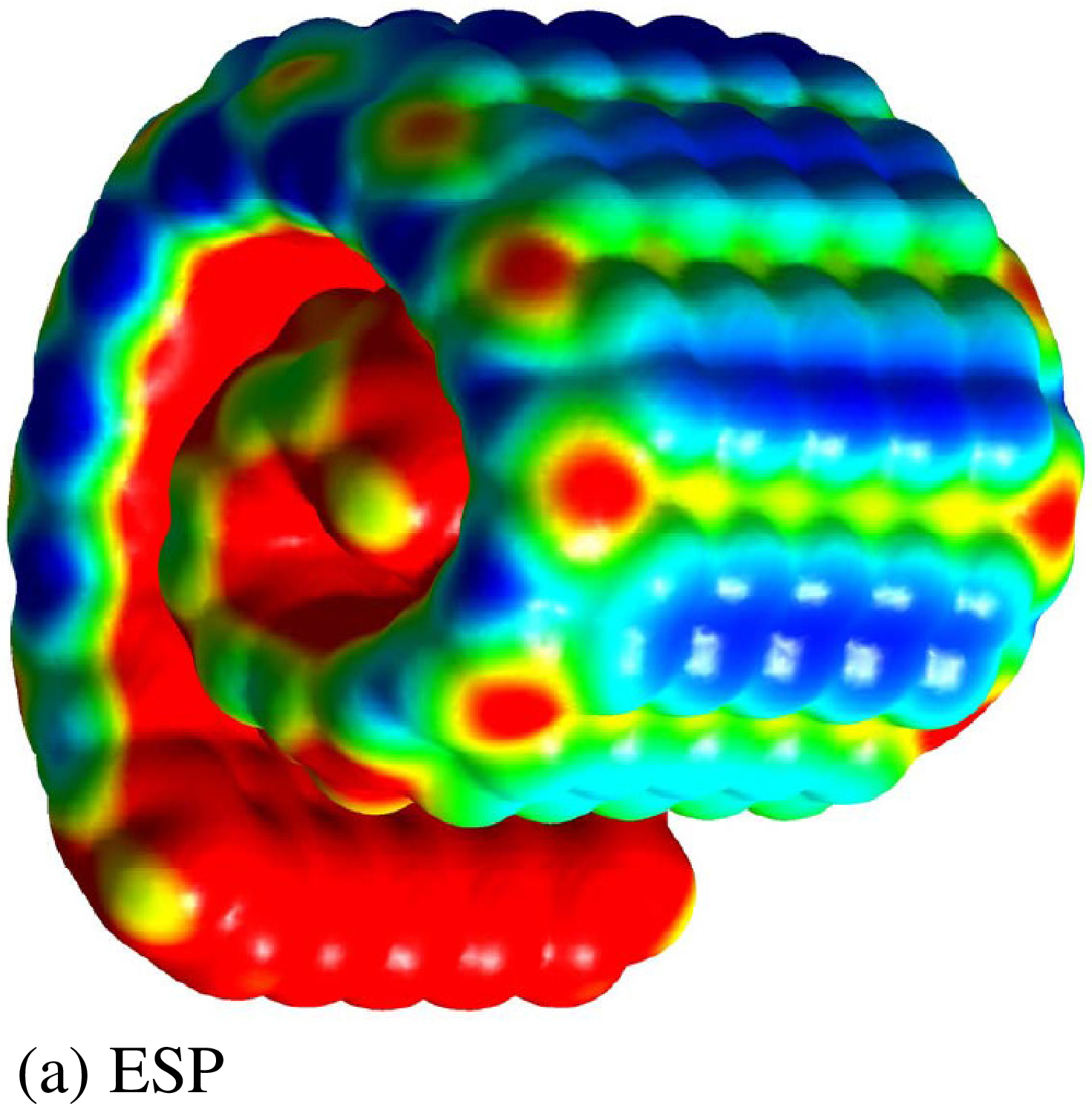}
\includegraphics[width=0.95\linewidth]{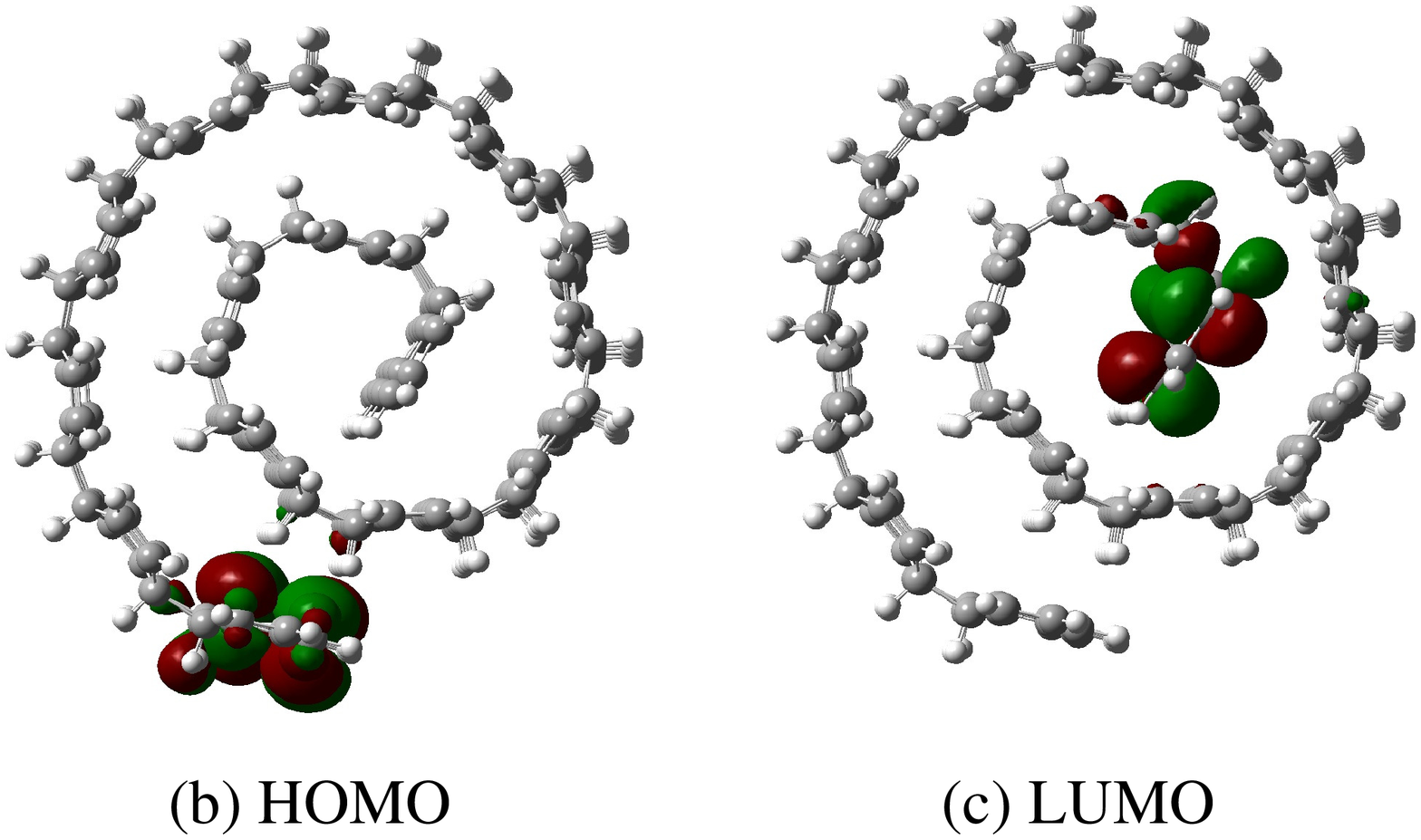}
\caption{(Color online)   (a) Electrostatic potential around
spiral graphone and the HOMO (b) and LUMO (c) of the system.
\label{figespROLL}}
\end{center}
\end{figure}

Figure~\ref{figespflat}(a) shows the electrostatic surface potential
(ESP) of B-GO before the completion of the rolling process. The
corresponding HOMO and LUMO are shown in Figs.~\ref{figespflat}(b,c)
respectively. During rolling both HOMO and LUMO are localized at
both ends simultaneously which is a consequence of the equivalent
geometry. The ESP of completely rolled B-GO is shown in
Fig.~\ref{figespROLL}(a). The ESP indicates that the spiral
structure has ionic characteristics, the red (blue) color indicating
positively (negatively) charged regions, which stabilizes the spiral
geometry even more.

\begin{figure}
\begin{center}
\includegraphics[width=0.95\linewidth]{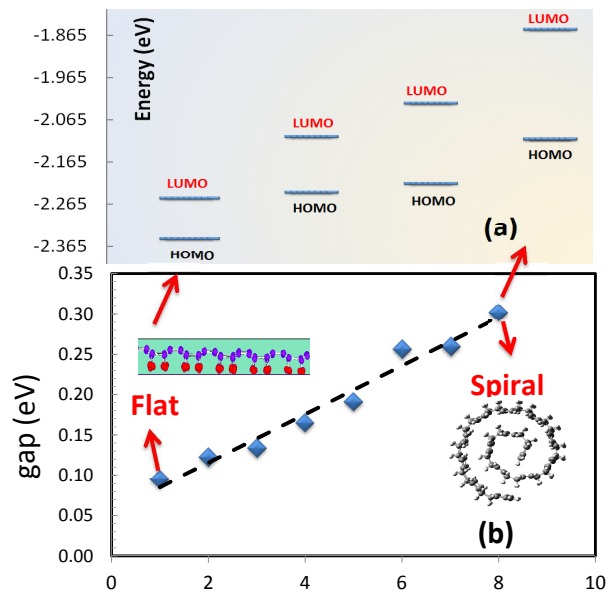}
\caption{(Color online)   (a) HOMO-LUMO energy diagram for four
different instances of the rolling process and (b) the variation of
the energy gap during rolling (for eight steps) for the B-GO shown
in Fig.~\ref{figDFT}(left). \label{figgap}}
\end{center}
\end{figure}

Surprisingly, after complete rolling, the HOMO and the LUMO are
separated and both localized in only one end, as can be seen from
Figs.~\ref{figespROLL}(b-c), whereas before complete rolling they
were localized in both ends simultaneously as already mentioned
before (see Figs.~\ref{figespflat}(b,c)). These effects are related
to the differences in electronegativity of the simulated atoms
(H$<$C$<$F). Because of the difference in ESP at the two edges, the
increment of the energy of the LUMO is larger than that of the HOMO,
which results in a widening of the energy gap from 0.1\,eV for the
flat sheet to 0.3\,eV for the spiral structure. This is shown more
explicitly in Fig.~\ref{figgap}, which reveals an almost linear
increase of the gap during the evolution to a spiral.

\subsection{Ab-initio results: One sided fluorinated graphene}
\begin{figure}
\begin{center}
\includegraphics[width=0.49\linewidth]{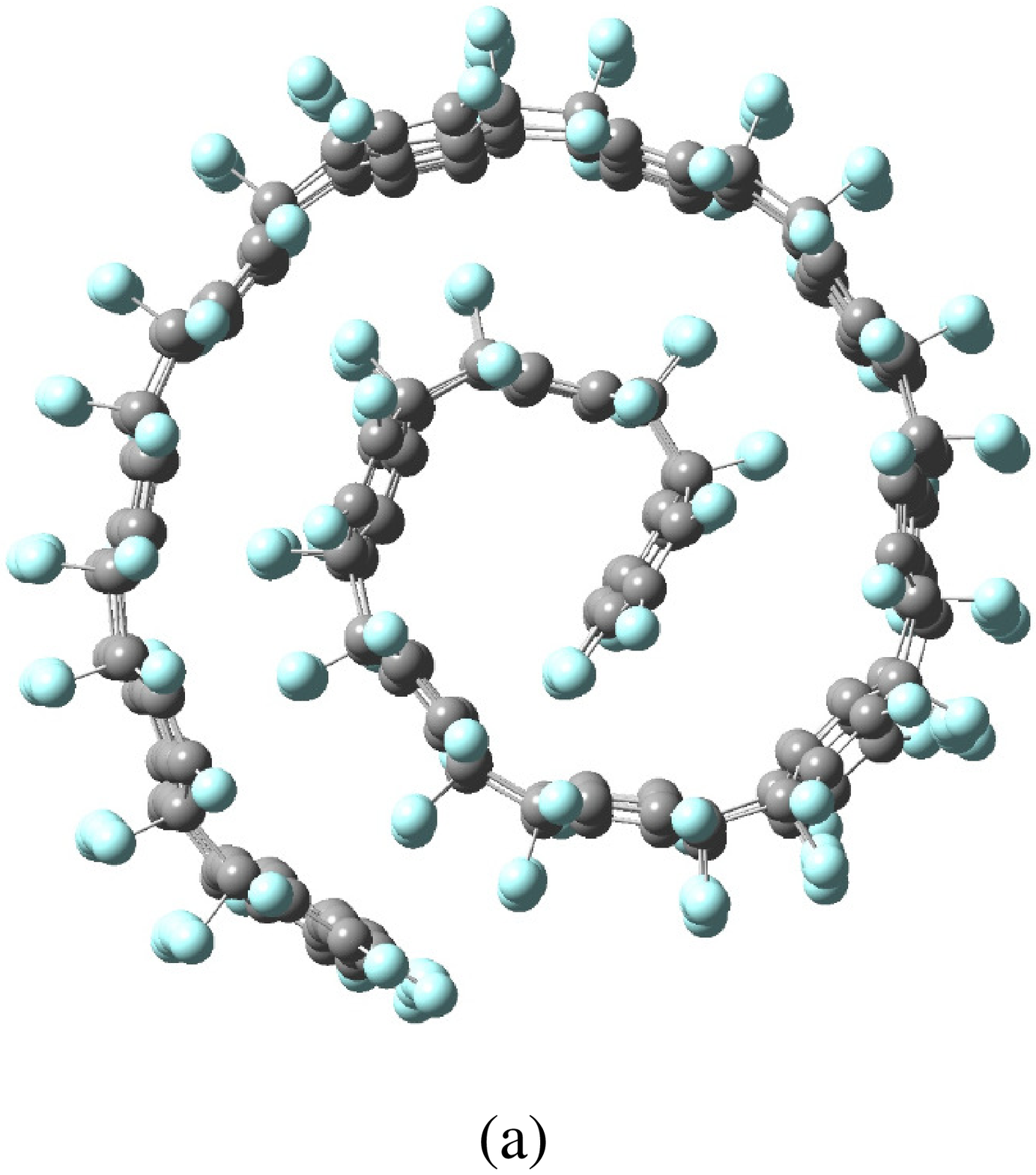}
\includegraphics[width=0.49\linewidth]{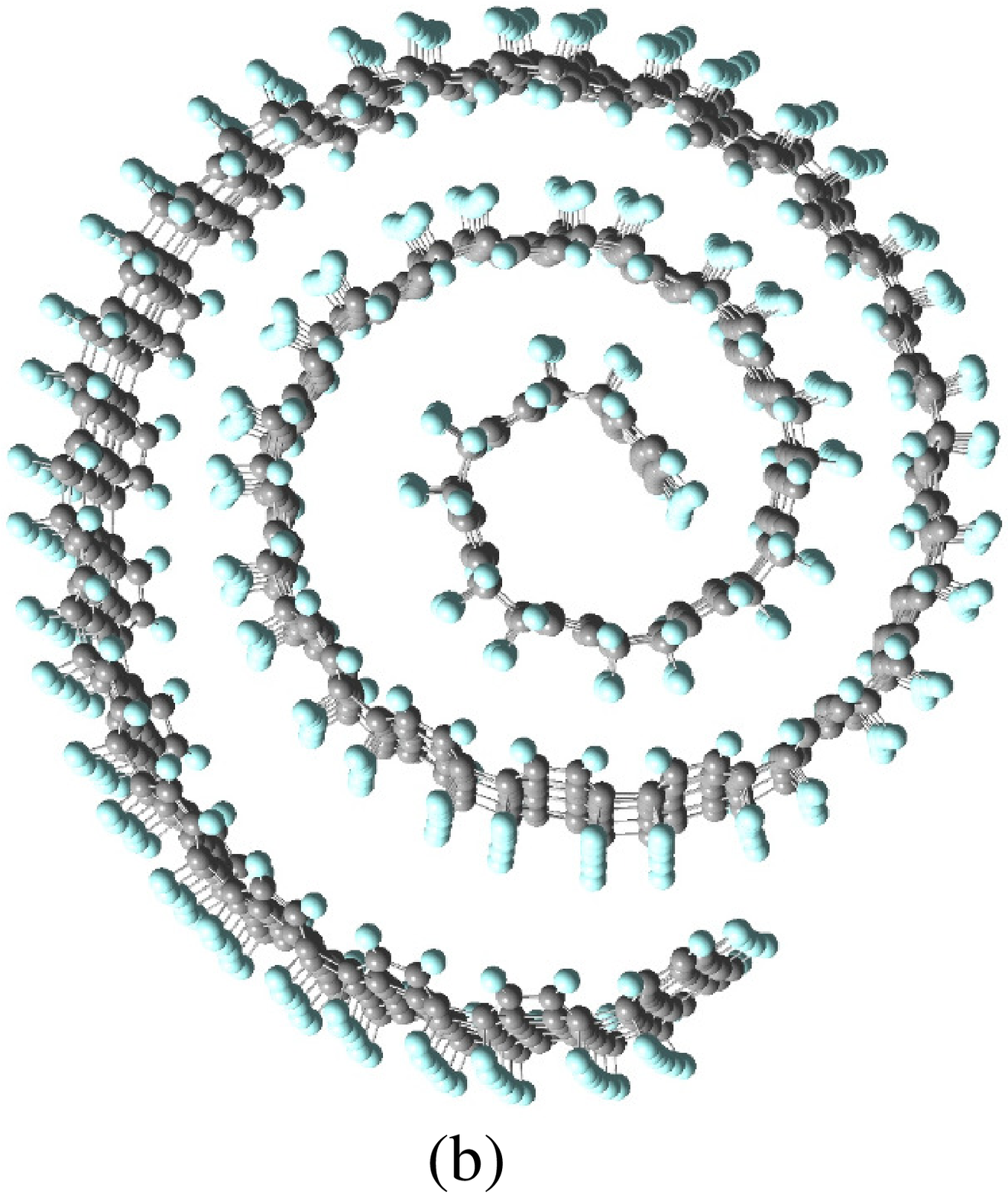}
\caption{(Color online)
Spiral B-GF from DFT calculations for 504 atoms (a).
Spiral B-GF from semiempirical calculations (PM6) for 1428 atoms (b).
\label{figGF}}
\end{center}
\end{figure}

Very similar results are found for GFs. Figure~\ref{figGF}(a) shows
the spiral GFs obtained using B3LYP/6-31G*. The spiral GF formed using
the PM6 method is shown in Fig.~\ref{figGF}(b). These configurations
are minimum energy configurations of B-GF. Similarly as for GO, the
rolling is due to the one-sided, partial (here 50 \%) coverage with
F atoms and corresponding sp$^3$ hybridization. The rolling process
starts from free boundaries where there is no resistance against any
torque.

\begin{figure}
\begin{center}
\includegraphics[width=0.6\linewidth]{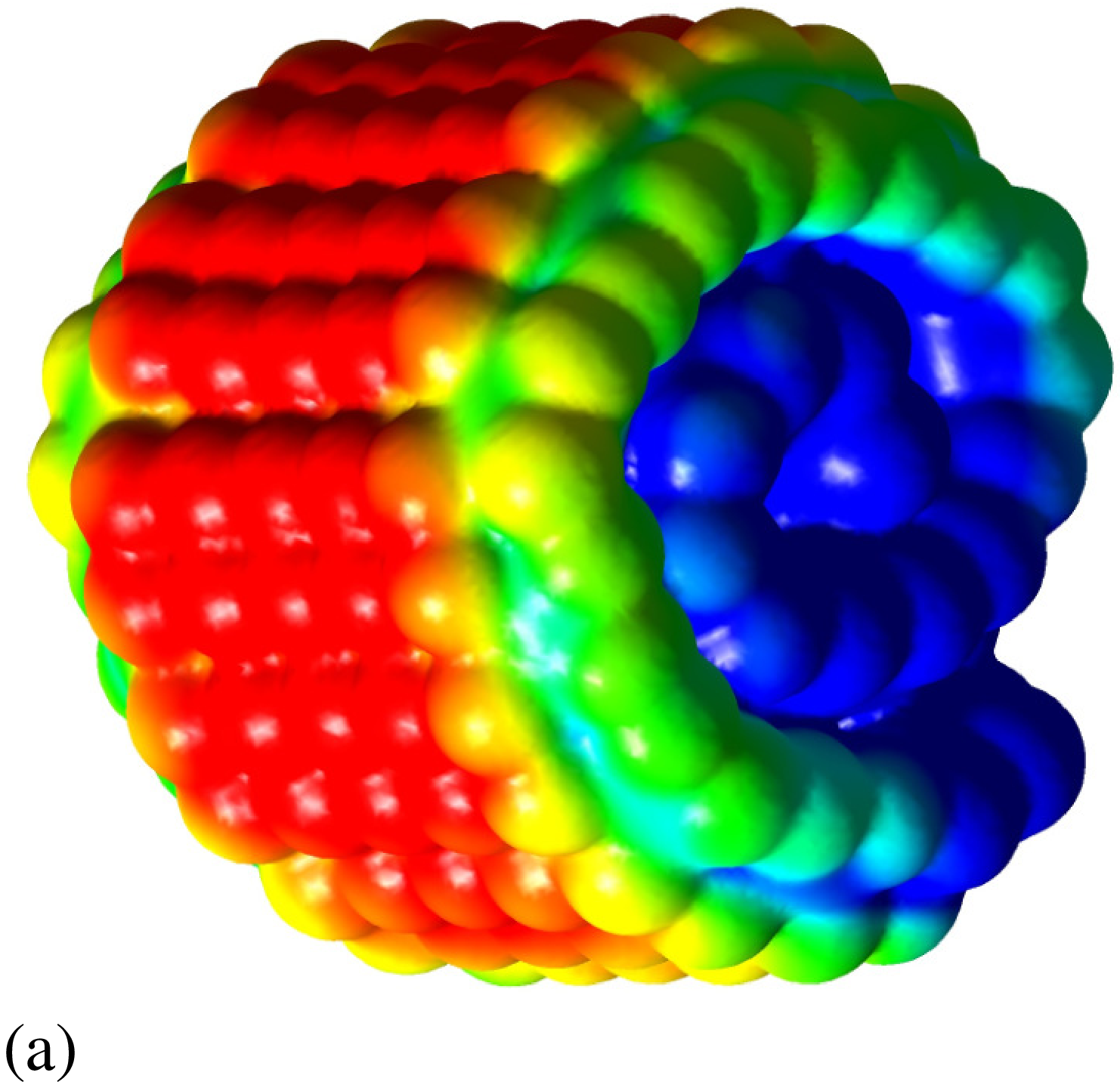}
\includegraphics[width=0.95\linewidth]{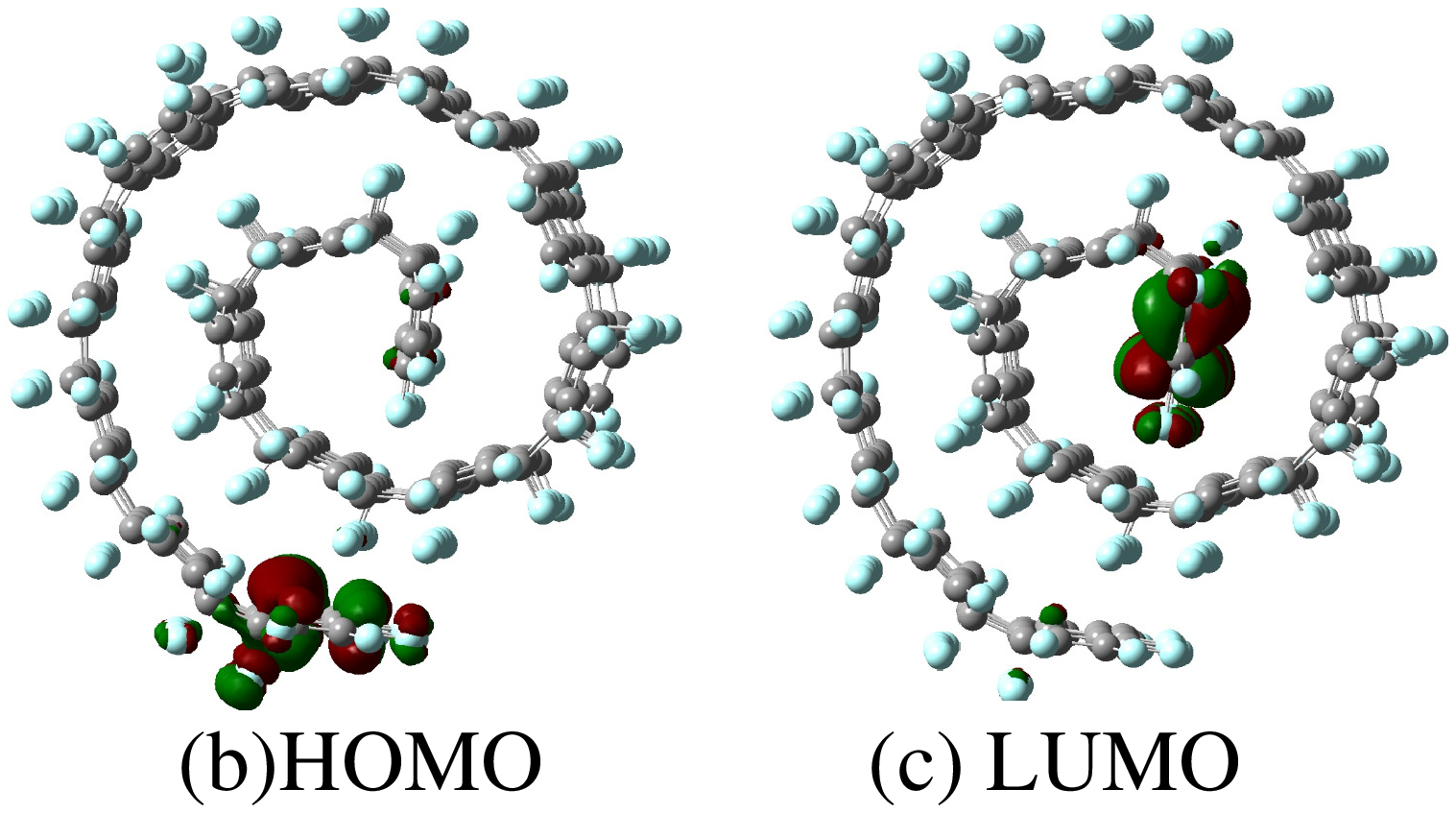}
\caption{(Color online)   ESP(a), HOMO (b) and LUMO (c) for the B-GF
system shown in Fig.~\ref{figGF}(left) \label{figesp}}
\end{center}
\end{figure}

%\subsection{One sided hydrogenated carbon nano-tube}
%One naturally expects that the one side-hydrogenated carbon nanotube
%(CNT) can be created by GO by manipulating the appropriative
%boundary conditions. In order to create hydrogenated CNT we applied
%periodic boundary condition along the zig-zag edge of an AC-GO
%nano-ribbon having width $L$. After relaxing the system at 10\,K we
%found that always a (5,5) CNT is formed independent of the
%longitudinal length. The width for B-GO, i.e. L=13.4~\AA~ agrees
%with the perimeter of a (5,5) CNT i.e. $5\sqrt{3}a_0$ where we take
%$a_0\cong1.5$\AA.~Fig.~\ref{CNT} shows the time evolution for the
%rolling of an AC-GO which eventually forms a (5,5) hydrogenated CNT.
%This is  consistent with previous results on the formation of CNT
%from partially hydrogenated graphene~\cite{nanolett2007}. Notice
%that in Fig.~\ref{CNT}  we removed hydrogens from the two lateral
%ends to avoid  possible C-H bonding when the sheet is rolled up to
%form a tube. It is important to note that the formation of CNT when
%taking into account the vdW force is impossible because in the last
%rolling step the vdW repulsion prevent the bonding creation between
%the two closed ends. Moreover, note that the B-GO nano-ribbon can
%also be deformed to CNT which is not the main objective of this
%paper.

\subsection{Classical molecular dynamics simulation results: large size flakes}

Using classical simulations, we first investigated the possibility
to create hydrogenated CNT from one-sided hydrogenated AC-GO
nano-ribbons, with periodic boundary condition along the zig-zag
edge, having a specific length equal to $L$=13.4~\AA~ between the
two free ends (in the flat geometry) . After relaxing the system at
10\,K we found that always a (5,5) CNT is formed independent of the
dimension of the box in the periodic direction (see movie in
supplementary material~\cite{suppl}). The used length for B-GO, i.e.
$L$=13.4~\AA, agrees with the perimeter of a one-sided hydrogenated
(5,5) CNT, i.e. $5\sqrt{3}a_0$ with $a_0\cong1.5$\AA.~The movie in
Ref. ~\cite{suppl} shows the time evolution for the rolling of an
AC-GO which eventually forms a (5,5) hydrogenated CNT. This is
consistent with previous results on the formation of CNT from
partially hydrogenated graphene~\cite{nanolett2007}. Note however
that the formation of these nanotubes is constrained by the choice
of the length of the ribbon, which has to be relatively small and
should match with the size of an (n,n) tube.

\begin{figure*}
\begin{center}
\includegraphics[width=0.535\linewidth]{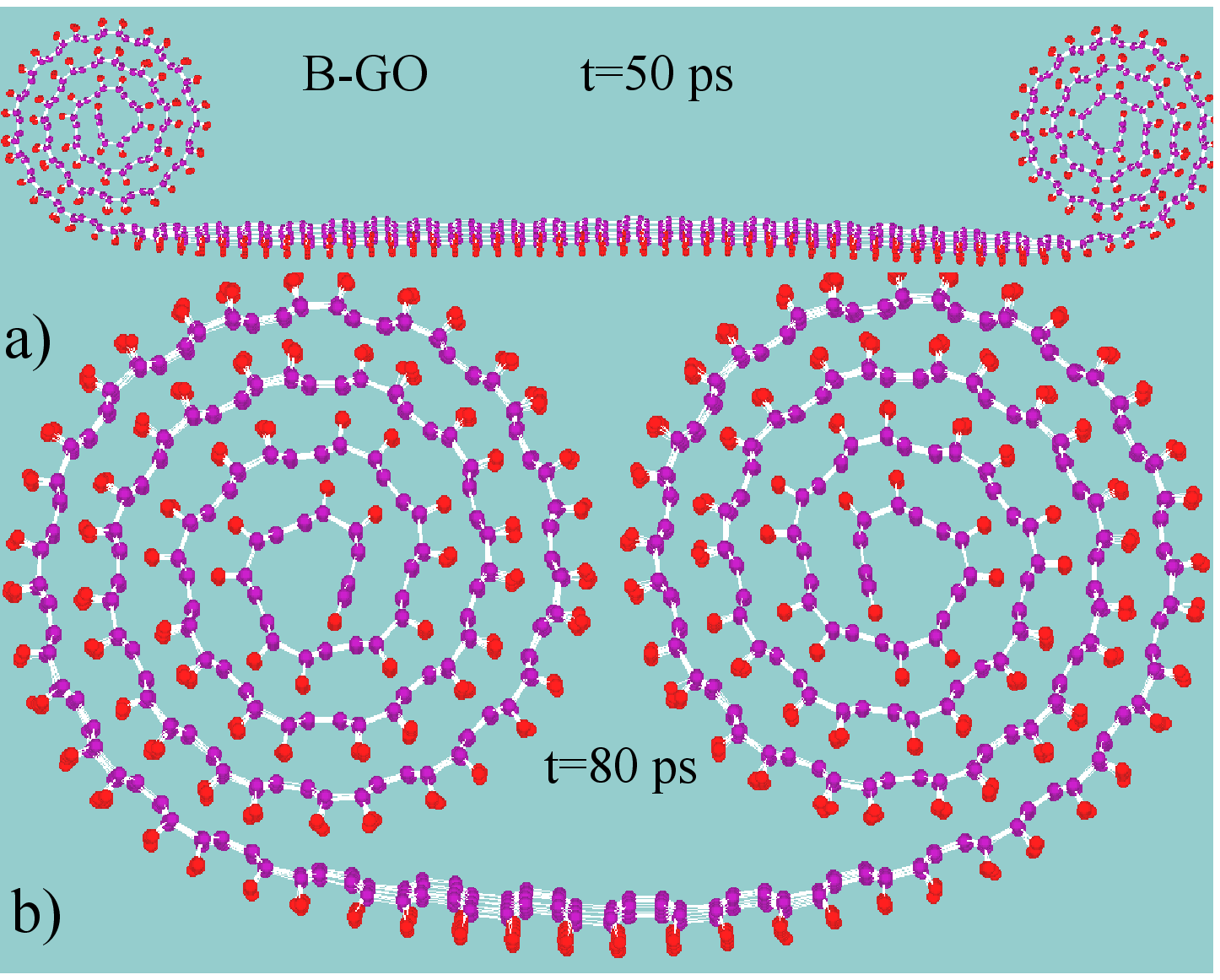}
\includegraphics[width=0.41\linewidth]{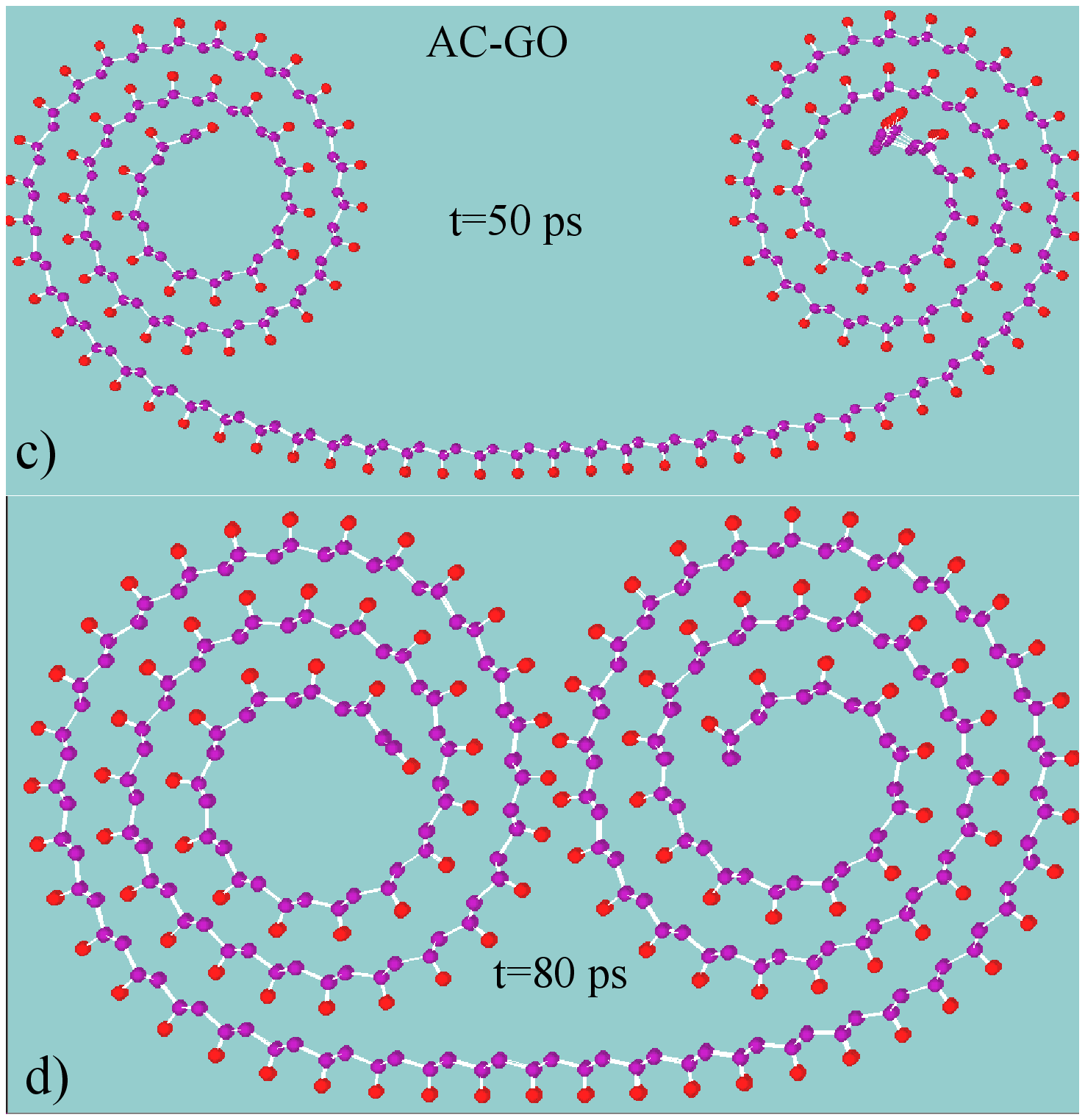}
\caption{(Color online) The rolled boat (arm-chair) like graphone
nano-ribbons after t=50\,ps (a,c) and 80\,ps (b,d). These systems
are stable up to at least 1000\,K. \label{fig1}}
\end{center}
\end{figure*}

In the remainder of this section we show the rolling effect for much
longer nano-ribbons, not leading to tube formation, using classical
MD simulation. We used graphone nano-ribbons with free zig-zag edges
and periodic boundary conditions in the direction parallel to these
edges (lateral side) and found that the final, minimal energy
configuration of both GO and GF is a spiral structure
similar to what we found with DFT. Figures~\ref{fig1}(a,b)
show two snap shots from the rolled B-GO which are free at two
longitudinal ends (here set to be zig-zag edges).
Similar results are found for AC-GOs, see
Figs.~\ref{fig1}(c,d). As expected, the corrugations in B-GO are
larger than in AC-GO. Thus, unrolling B-GO costs much more energy
than AC-GO.

\begin{figure}
\begin{center}
\includegraphics[width=0.9\linewidth]{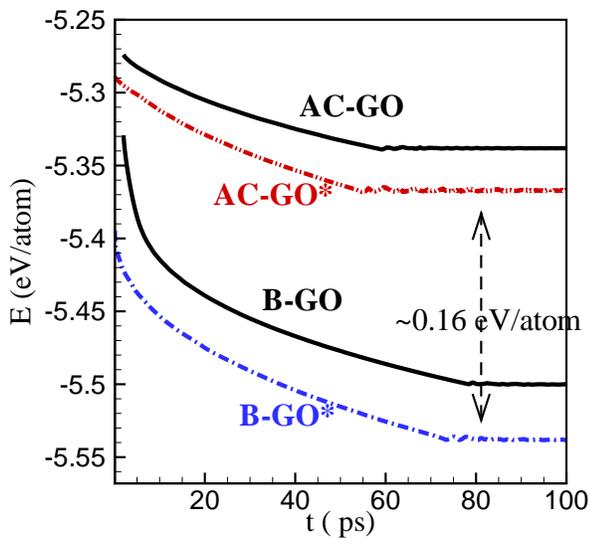}
\caption{(Color online)   Variation of the total energy with time
for both arm-chair and boat-like graphone. \label{fig3}}
\end{center}
\end{figure}

Figure~\ref{fig3} shows the variation of the total energy per atom
as a function of the rolling process time for AC-GO, B-GO, AC-GO*
and B-GO*. Clearly B-GO* is the most stable having the lowest
energy. The energy gap between AC-GO (AC-GO*) and B-GO (B-GO*) is
about 0.16 eV/atom. Furthermore it is seen that including the vdW
energy term lowers slightly the energy by $\sim0.04$ eV/atom due to
the attractive van der Waals interaction. Notice that the
ionic characteristic of the rolled structure cannot be reproduced by
these classical simulation with AIREBO. Nonetheless, the classically
found spiral structures are quite stable, which confirms that the
main driving force for the rolling is the one-sided absorption and
corresponding sp$^3$ hybridization and not the ionic characteristics
found in the DFT calculations.

\begin{figure}
\begin{center}
\includegraphics[width=0.9\linewidth]{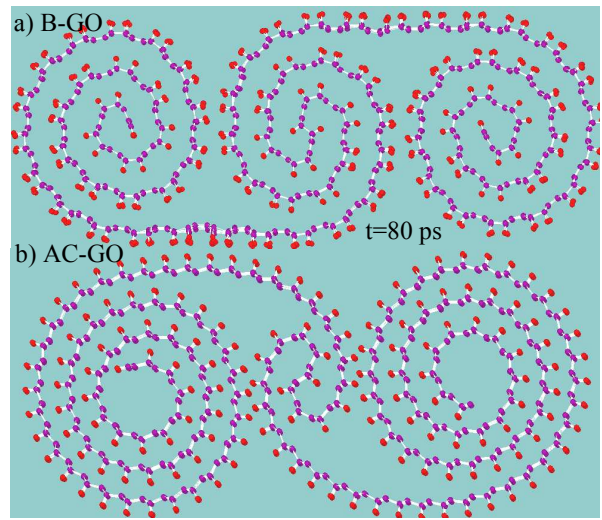} \caption{(Color
online) The optimized configuration of a boat-like (a) and arm-chair
like (b) semi-infinite graphone. in (a) the A-lattice and in (b) the
B-lattice of carbon sites were covered by hydrogens. \label{fig4}}
\end{center}
\end{figure}

In Fig.~\ref{fig4} rolled configurations of a system consisting of
two connected GO parts that are hydrogenated at opposite sides are
shown. Hydrogens at $x<$0 ($x>$0) are bonded to the upper (lower)
side of the sheet. Notice that the AC-GO and B-GO form different
patterns of rolled GO with an extra rolled part in the middle. This
hints to the interesting possibility  to engineer the rolled shape
of GO carefully by selectively hydrogenating parts of the initial
graphene sheet.

%\begin{figure}
%\begin{center}
%\includegraphics[width=0.98\linewidth]{ReaxF-F.eps}
%The rolled B-GF using molecular dynamics simulations based on ReaxFF potential.
%
%\caption{(Color online) \label{figGFMD}}
%\end{center}
%\end{figure}
%Very similar deformations were found for GFs using the ReaxFF
%potential in molecular dynamics simulations. Figure~\ref{figGFMD}
%shows the rolled B-GF after 40 ps.

\begin{figure}
\begin{center}
\includegraphics[width=1\linewidth]{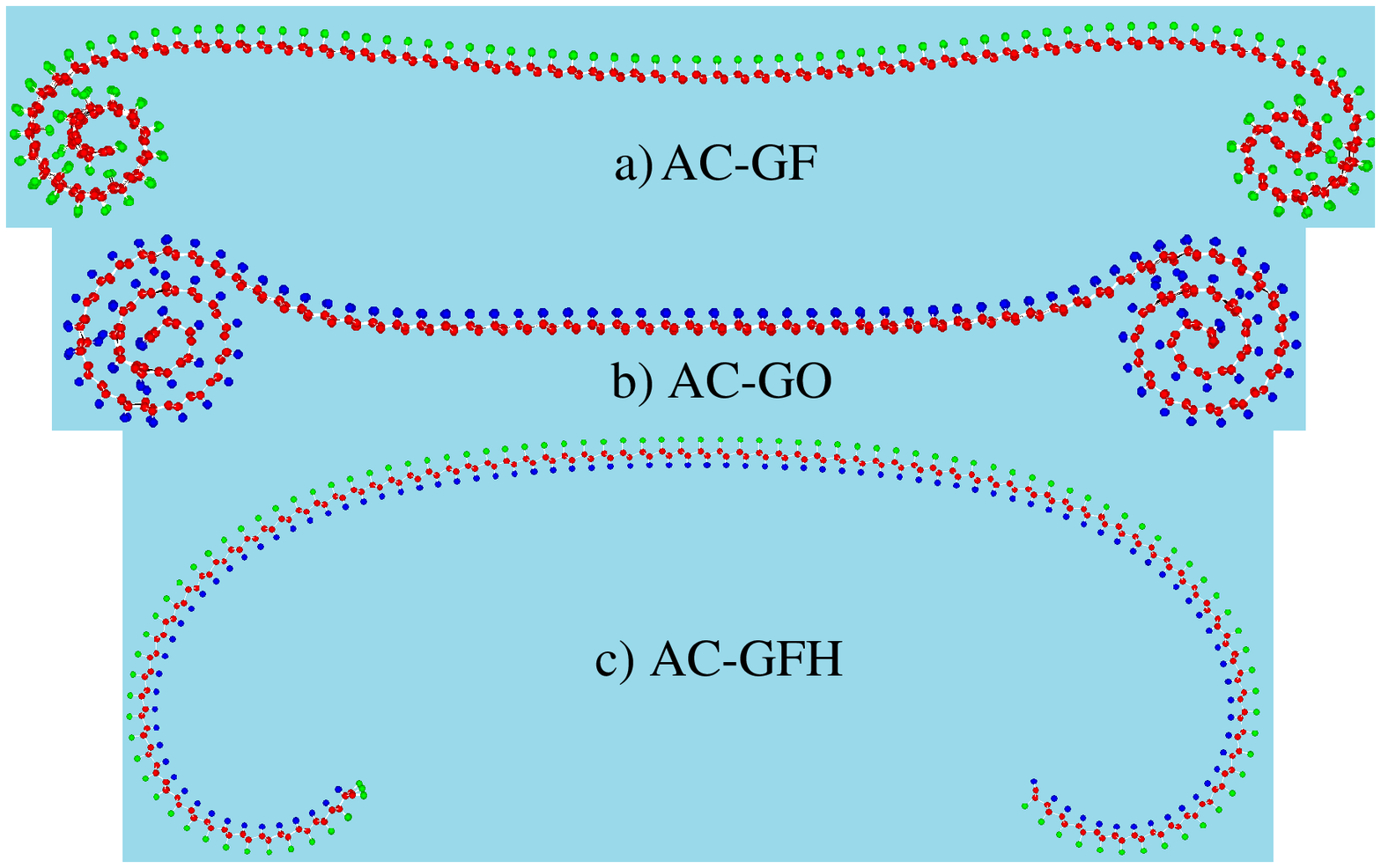}
\caption{(Color online)  \textbf{Side view of three different spiral
structures: AC-GF (a), AC-GO(b) and AC-GFH(c) after 25 ps of
relaxation from the starting flat geometry.} \label{figGFH}}
\end{center}
\end{figure}

As an additional, final result of this section we present the
rolling behavior of a system with one side covered by hydrogen and
the another side covered by fluor, which we called `GFH'. Recently,
by using ab-initio calculations, a negative formation energy was
found for this system~\cite{kast2010}. In Fig.~\ref{figGFH} side
views are shown for the rolled structure of AC-GF (a), AC-GO (b) and
AC-GFH (c) after 25 ps of molecular dynamics simulations (red, green
and blue balls are carbon, fluor and hydrogen, respectively). In all
three cases we used the ReaxFF potential and the simulations were
done at T=10\,K, as before. It is clear from Fig.~\ref{figGFH} that
AC-GO (AC-GFH) has the largest (smallest) curvature, due to the fact
that when one side is covered by hydrogen and the other side is
covered by fluor, the driven force for rolling is partially
balanced, but not completely as the fluor atoms are larger leading
to a stronger intra-layer repulsion at the fluor side, so that the
system still tends to roll, albeit with a much smaller curvature.

\subsection{Temperature effects}
New structures have to be tested in terms of their stability against
temperature. In order to study the temperature effects on the spiral
B-GO, after obtaining it at 10\,K (note that in principle spiral GO
can also be formed at room temperature but we set $T$=10\,K in order
to minimize thermal fluctuations and to find more rapidly the ground
state spiral configurations) we raised the temperature up to
1000\,K. Not so surprising in view of the significant gain in
binding energy of ~0.1 eV/atom (~1200 K) for AC-GO and much more for
B-GO (see Fig. 9) in their rolled states with respect to the flat
configurations, the spiral GOs keep their configuration. Beyond
700\,K the loops start to deform slightly. However, even at 1000\,K
the system is not un-zipped and the spiral configuration is
conserved~\cite{suppl}  at least within the here used MD simulation
time of 1~ns. For T$>$1000\,K the system starts to separate into two
different loops~\cite{suppl}. From these simulations we conclude
that spiral GO should be rather stable at room temperature,
facilitating its possible synthesis in future experiments. Two
movies are provided in the supplementary materials~\cite{suppl} to
show the rolling of B-GO at T=10\,K and to show the stability of
B-GO during the heating process up to T=1000 \,K. For GFs the spiral
system is somewhat less stable. Close to 800 K the GFs start to
un-roll and beyond this temperature they start to be fractured.
Notice that the C-H bonds (C-F bonds) in GOs (GFs) are stable even
after unrolling, thus no hydrogen release was observed when heating
up the rolled sheets within our simulation time of 0.1\,ns.

We also applied external uniaxial
compression along the axis of the spiral structures in order to
investigate the mechanical stability of these structures, which
turned out to resists against the external pressure much better than
graphene. Detailed mechanical properties of these new spiral
hydrocarbons need further investigations.

\section{Conclusions}
Free boat like and arm-chair like one-sided hydrogenated/fluorinated
graphene (graphone/fluorographene) ribbons/flakes  spontaneously
roll up to form spiral configurations which should be quite stable
at room temperature. The spiral structures exhibit strong mechanical
rigidity preventing them to unroll. The main driving force behind
the spiral formation is the energy relaxation associated with the
one-sided, asymmetric orientation of sp$^3$ bonding. Further
modestly stabilizing factors are the van der Waals attractive
interaction between the stacked shells and (only in the DFT
calculations) the ionic interactions between the shells, which
appear to be polarized across the shell width. Boat-like
hydrogenated/fluorinated graphene yields more stable, spiral systems
than arm-chair like graphone. \textbf{The graphene sheet where one
side is covered by hydrogen and the other side by fluor is unstable
and it also forms a rolled up structure, albeit with a smaller
curvature.}. The highest occupied and lowest unoccupied orbital for
spiral GO and GF are localized at opposite ends of the system. The
energy gap increases when the
system evolves from the flat to the spiral shape.\\

%%%%%%%%%%%%%%%%%%%%%%%%%%%%%%%%%%%%%%%%%%%%%%%%%%%%%%%%%
\vspace{0.1cm} \textbf{\emph{{Acknowledgments}}}\\ We thank A.
Sadeghi, M. R. Ejtehadi and J. Amini for their useful comments. This
work is supported by the ESF EuroGRAPHENE project CONGRAN and the
Flemish Science Foundation (FWO-Vl).
%%%%%%%%%%%%%%%%%%%%%%%%%%%%%%%%%%%%%%%%%%%%%%%%%%%%%%%%%%%

%\bibitem{expF2}J. T. Robinson, J. S. Burgess, C. E. Junkermeier, S. C. Badescu, T. L. Reinecke, F. K. Perkins, M. K. Zalalutdniov, J. W. Baldwin, J. %C. Culbertson, P. E. Sheehan, and E. S. Snow, Nano Lett. {\bf10}, 3001 (2010).

%%%%%%%%%%%%%%%%%%%%%%%%%%%%%%%%%%%%%%%%%%%%%%%%%%%%%%%%%%%

\end{document}